\documentclass{elsart} \usepackage{graphicx}
\begin{document}
\begin{frontmatter}
\title{Application of the Time of Flight Technique for Lifetime
Measurements with Relativistic Beams of Heavy Nuclei}

\author{A. Chester,} \author{P. Adrich,} \author{A. Becerril,}
\author{D. Bazin,} \author{C. M. Campbell,} \author{J. M. Cook,}
\author{D.-C. Dinca,\thanksref{labelDCD}} \author{W.F. Mueller,}
\author{D. Miller,} \author{V. Moeller} \author{R. P. Norris,}
\author{M. Portillo,} \author{K. Starosta,\corauthref{cor}}
\author{A. Stolz,} \author{J. R. Terry} \author{H. Zwahlen,}
\author{C. Vaman}

\address{National Superconducting Cyclotron Laboratory and\\
Department of Physics and Astronomy, Michigan State University,\\ 164
S. Shaw Lane, East Lansing, Michigan 48825-1321}

\thanks[labelDCD]{present address: American Science \& Engineering,
Inc. 829 Middlesex Turnpike Billerica MA 01821}

\corauth[cor]{corresponding author: phone (517) 324-8138, fax: (517)
353-5967, e-mail: starosta@nscl.msu.edu}

\collab{A. Dewald}

\address{Institute for Nuclear Physics, University of Cologne,
Z{\"u}lpicher Str. 77, D-50937 K{\"o}ln, Germany }

\begin{abstract}

A novel method for picosecond lifetime measurements of excited
$\gamma$-ray emitting nuclear states has been developed for fast beams
from fragmentation reactions. A test measurement was carried out with
a beam of $^{124}$Xe at an energy of $\sim$55 MeV/u. The beam ions
were Coulomb excited to the $2^+_1$ state on a movable target. Excited
nuclei emerged from the target and decayed in flight after a distance
related to the lifetime. A stationary degrader positioned downstream
with respect to the target was used to further reduce the velocity of
the excited nuclei. As a consequence, the $\gamma$-ray decays from the
$2^+_1$ excited state that occurred before or after traversing the
degrader were measured at a different Doppler shift.  The $\gamma$-ray
spectra were analyzed from the forward ring of the Segmented Germanium
Array; this ring positioned at 37$^\circ$ simultaneously provides the
largest sensitivity to changes in $\beta$ and the best energy
resolution. The ratio of intensities in the peaks at different Doppler
shifts gives information about the lifetime if the velocity $\beta$ is
measured.  The results and range of the application of the method are
discussed.

\end{abstract}
\end{frontmatter}

\section{Motivation}

Electromagnetic transition matrix elements are vitally important for
nuclear structure studies. These matrix elements can be extracted if
the information on the transition rate or lifetime is available for
the state of interest.  Picosecond-range lifetimes of excited nuclear
states which decay by $\gamma$-ray emission can be reliably measured
using Doppler shift methods; in particular, the Recoil Distance
Method, also known as the plunger technique, see for example Refs.\
\cite{Fos74,Kru00}.

Exciting research opportunities to study nuclei far from stability
have opened recently at the National Superconducting Cyclotron
Laboratory following the construction of the Coupled Cyclotron
Facility\ \cite{Mil01}.  These nuclei are produced in fragmentation
reactions at primary beams energies of $\sim$140 MeV/u\ \cite{Sto05}.
The fragments of interest are selected and identified on an event by
event basis using the A1900 magnetic spectrograph\ \cite{Mor03}. The
relativistic energies of secondary beams result in significant Doppler
shifts for $\gamma$-rays following the secondary reactions. In fact,
the Doppler broadening of $\gamma$-ray peaks resulting from momentum
distribution of the secondary reaction products as well as from
$\gamma$-ray detector opening angles is known to be the limiting
factor in the application of high-resolution spectroscopy in
relativistic beam facilities\ \cite{Gla98}. The current paper reports
on a method of utilizing these large Doppler shifts for picosecond
lifetime measurements using a relativistic plunger technique.

\section{Experimental Method}	

In the reported experiment a beam of $^{124}$Xe at an initial energy
of 140 MeV/u was provided by the Coupled Cyclotron Facility at the
National Superconducting Cyclotron Laboratory \ \cite{Mil01}. The
primary beam was degraded in the fragmentation target of the A1900
separator\ \cite{Mor03} and delivered at an average energy of 54.9
MeV/u to the plunger apparatus installed at the target position of the
S800 spectrometer\ \cite{Baz03}. The A1900 momentum acceptance cut was
set to 0.5\% and it was estimated that the plastic scintillator
detectors used downstream from the A1900 contributed less than 0.02\%
to the beam momentum spread.

In the plunger apparatus the beam ions were Coulomb excited to the
2$^+_1$ state on a movable $^{93}$Nb target with a thickness of 91.5
mg/cm$^2$. Excited projectiles emerged from the target with velocity
$\beta_H$=0.278(1) and $\gamma$-decayed in flight after a distance
governed by the lifetime of the $^{124}$Xe first excited state. A
stationary 79.5 mg/cm$^2$ $^{27}$Al degrader positioned downstream
with respect to the target was used to further reduce the velocity of
these excited nuclei to $\beta_L$=0.178(1). The $\gamma$-ray intensity
measurements were carried out using the Segmented Germanium Array
SeGA\ \cite{Mue01} centered on the degrader foil. Four target-degrader
distances of $d$=15.8 mm, 10.5 mm, 5.3 mm and a contact between target
and degrader foils at $d$=0 mm were investigated. The separation
between target and the degrader was fixed using thin cylindrical
aluminum spacers and the precision of the spacer length measurement
was better than 0.1 mm.

\begin{figure}[h]
\begin{center}
\includegraphics[clip=true,angle=-90, scale=0.50]{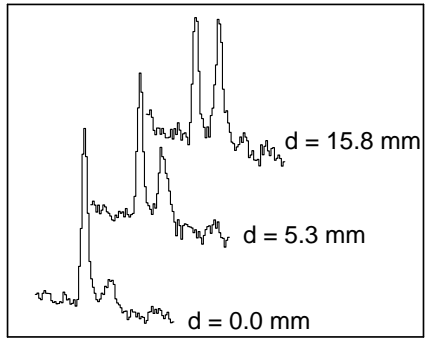}
\caption{Change of intensity as a function of the target-degrader
distance for the two components of a Doppler-shifted $\gamma$-ray peak
corresponding to the 354~keV transition from the first excited 2$^+_1$
state in $^{124}$Xe. The higher- and lower-energy component of the
peak corresponds to the decays upstream and downstream with respect to
the plunger degrader foil, respectively. The variation of intensities
of these two components as a function of the target/degrader
separation is defined by the lifetime of the transition and the
velocity $\beta_H$ of the nuclei between the target and the
degrader. The method reported in the current paper utilizes this
variation for lifetime measurement of the 2$^+_1$ state in
$^{124}$Xe.}
\label{fig1}
\end{center}
\end{figure}

Depending on whether the $\gamma$-ray decay occurred in flight between
target and degrader or if the decay occurred after slowing down in the
degrader, the $\gamma$-rays exhibited different Doppler shifts. The
resulting $\gamma$-ray spectra shown in Fig.\ \ref{fig1} exhibit two
peaks; the higher-energy peak corresponds to the $\gamma$-rays emitted
before traversing the degrader while the lower-energy peak corresponds
to $\gamma$-rays emitted after the degrader. From the relative
intensity of the peak areas as a function of target-degrader distance,
the lifetime of the state can be inferred as discussed in Sec.\
\ref{lifetimeSec}.

After passing through the degrader the nuclei were analyzed in the
S800 spectrometer.  Coincidence events corresponding to a $\gamma$-ray
detection in the SeGA array and a beam particle detection in the E1
plastic scintillator detector at the S800 focal plane\ \cite{Yur99}
were recorded for off-line analysis. In addition, the number of
particles detected by the S800 E1 plastic scintillator was monitored
and stored in intervals of five-second; these data were used for
normalization of $\gamma$-ray intensities measured at different
target-degrader separations.  The readouts from the S800 focal plane
Ionization Chamber were present in the data stream and were used in
the off-line analysis for particle identification discussed in Sec.\
\ref{PID}. The average rates for the experiment were $\sim$23000
particles per second at the S800 focal plane and $\sim$1700 particles
per second in SeGA with the SeGA/S800 coincidence rate of $\sim$60 per
second. The average live time of the data acquisition system was
$\sim$98\% and constant for runs at different target-degrader
separation.

An important difference between the plunger technique applications at
intermediate energies of $\sim$55 MeV/u reported in this paper and the
analogous technique near the Coulomb barrier energies of $\sim$5 MeV/u
is a significant contribution from the reactions on the degrader to
the measured $\gamma$-ray intensity. In the current work the
experimental information on that contribution has been identified from
a separate run with the $^{27}$Al plunger degrader foil in place but
with the excitation target removed from the field of view of the SeGA
detectors.  To assure that the beam energy in this run is matched to
that of the runs at smaller target-degrader separations the beam
passed through the 91.5 mg/cm$^2$ $^{93}$Nb foil mounted $\sim$ 60 cm
upstream of the plunger degrader. From these data it has been verified
that the $\gamma$-rays following the reactions on the degrader are
emitted at the Doppler shift corresponding to the velocity $\beta_L$
and thus are indistinguishable from the $\gamma$-ray decays from
excitation on the plunger target which are emitted after traversing
the degrader. The impact of the reactions on the degrader on the
extracted lifetime is discussed further below.

\section{Lifetime Measurement with the Time of Flight Technique}
\label{lifetimeSec}
Coulomb excitation at the 55 MeV/u energy of interest for the current
experiment is known to be predominantly a single step process\
\cite{Win79}.  For a case of a single level without any feeding, the
number $N_H$ and $N_L$ of $\gamma$ rays detected for states decaying
between the target and the degrader and downstream from the degrader,
respectively, are given by the relations:
\begin{eqnarray}
\label{NHmaster}
N_H(d)&=&{\frac{I_0}{D_H}}\int_{-(d+d_0+d_m)}^{-d_m}\varepsilon_H(z)
\exp(-{z+d+d_0+d_m\over D_H}) dz\\
\label{NLmaster}
N_L(d)&=&{\frac{I_0}{D_L}}{\exp(-{d+d_0+d_d\over D_H})}
\int_{d_d-d_m}^{d_{max}}\varepsilon_L(z) \exp(-{z-d_d+d_m\over D_L})
dz
\end{eqnarray}
with $D_H$ and $D_L$ being related to the lifetime $\tau$ through:
\begin{equation}
D_H =\tau \mbox{c} {\beta_H \over \sqrt{1-\beta_H^2}}
\end{equation}
\begin{equation}
D_L =\tau \mbox{c} {\beta_L \over \sqrt{1-\beta_L^2}}
\end{equation}

\begin{figure}[h]
\begin{center}
\includegraphics[clip=true,angle=0, scale=0.50]{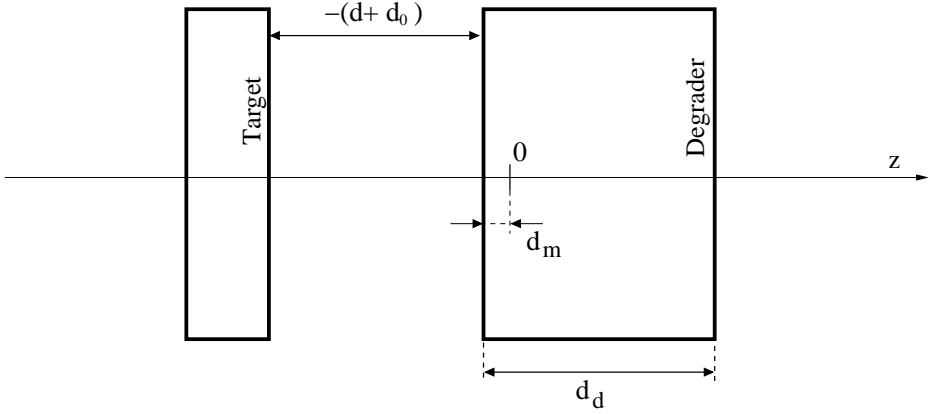}
\caption{Schematic alignment of the plunger excitation target and the
degrader foil with respect to the center of the SeGA array.  The
distance $d$ is the measured target/degrader separation, $d_0$ is the
offset between the measured and the true target/degrader separation,
$d_m$ is the length by which the plunger degrader foil is misaligned
with respect to the center of the $\gamma$-ray array, and $d_d$ is the
degrader thickness.}
\label{fig2}
\end{center}
\end{figure}

Above, $I_0$ stands for the number of states excited in the target,
while $\varepsilon_H (z)$ and $\varepsilon_L (z)$ are the efficiencies
for detection of $\gamma$ rays emitted at point $z$ for the decays
between the target and the degrader and downstream from the degrader,
respectively. For the distance parameters, $d$ is the measured
target/degrader separation, the $d_0$ is the offset between the
measured and the true target/degrader separation, the $d_m$ is the
length by which the plunger degrader foil is misaligned with respect
to the center of the $\gamma$-ray array, and $d_d$ is the degrader
thickness. The upper integration limit $\d_{max}$ in Eq.\
\ref{NLmaster} corresponds to the largest distance from the target at
which $\gamma$-ray decays can still be detected in the SeGA array. In
the current experiment this distance is significantly larger than
$D_H$ and $D_L$ and $\d_{max}=\infty$ can safely be assumed. The
symbols for the plunger target/degrader foil alignment are further
explained in Fig.\ \ref{fig2}.

In Eq.\ \ref{NLmaster} it has been assumed that the decays in the
degrader occur at the projectile velocity $\beta_H$. This
approximation is valid for the current experiment since the thickness
of the degrader $d_d\approx 0.3$ mm is significantly smaller than $D_L
\sim D_H\approx 5.5$ mm. For thick degraders, the relevant correction
has to be determined based on the energy loss of the beam nuclei in
the degrader; however, it should be noted, that such a correction is
not going to alter the dependence of $N_L$ on d, which is of interest
for the lifetime measurement.

In Sec.\ \ref{efficiency} the efficiencies $\varepsilon_H (z)$ and
$\varepsilon_L (z)$ simulated with the Monte Carlo GEANT4 code\
\cite{Ago03} are discussed; these simulations demonstrate that these
efficiencies can be well approximated with a second order polynomial
in $z$
\begin{equation}
\label{parabola}
\varepsilon(z)=\varepsilon_0+\varepsilon_1 z +\varepsilon_2 z^2
\end{equation}
In this case:

\begin{eqnarray}
\label{NH}
N_H(d)& = & I_0 \left[ \tilde{\varepsilon}_H(1-exp(-{d+d_0\over D_H}))
+\epsilon_H(d) \right]\\
\label{NL}
N_L(d)& = & I_0 \tilde{\varepsilon}_L \exp(-{d+d_0+d_d\over D_H})
\end{eqnarray}

with

\begin{eqnarray}
\tilde{\varepsilon}_H
&=&\varepsilon_H(D_H-d_m)+\varepsilon_{2\;H}D^2_H\\
\label{epsd}
\epsilon_H(d)&=&\varepsilon_H(D_H-d_m-d-d_0)-\varepsilon_H(D_H-d_m)\\
\tilde{\varepsilon}_L
&=&\varepsilon_L(D_L+d_d-d_m)+\varepsilon_{2\;L}D^2_L
\end{eqnarray}

If a $d$-independent contribution from reactions on the degrader $N_D$
is taken into account the final formula for $N_L$ reads:
\begin{eqnarray}
\label{lowfit}
N_L(d)& = & I_0 \tilde{\varepsilon}_L \exp(-{d+d_0+d_d\over D_H})+N_D
\end{eqnarray}
It should be pointed out that the $N_L(d)$ depends exponentially on
the target/degrader separation $d$, while there are $d$-dependent
efficiency corrections to the number of decays between the target and
the degrader $N_H$ as given by Eqs.\ \ref{NH} and \ \ref{epsd}.

The above formulae demonstrate that the number of decays upstream and
downstream from the degrader is defined by the target/degrader
separation $d$ and the velocity $\beta_H$ between the target and the
degrader. Thus, the lifetime of the state of interest can be extracted
from the measured intensity variation of $N_H$ and $N_L$ as a function
of $d$ if the velocity $\beta_H$ is measured at the same time. The
intensity variations of $N_H$ and $N_L$ as a function of $d$ can be
measured from the number of counts in the low- and high-energy
component of the peak corresponding to the transition of interest
normalized to the beam current integrated over the time of the run.
The measurement of these experimental observables provides the basis
for the current method of lifetime determination.

\section{Efficiency of $\gamma$-ray detection}
\label{efficiency}

It is argued in the previous section that the dependence of the
efficiency for $\gamma$-ray detection on the position of a decay along
the beam axis is of importance for the application of the current
method.  Since relativistic effects have a significant impact on the
$\gamma$-ray detection efficiency, this dependence has been studied
for the purpose of the current experiment using a Monte Carlo GEANT4\
\cite{Ago03} simulation code. The code was set up to include major
parts of the plunger apparatus used in the experiment and provided
segment energy deposits for detected $\gamma$-rays in a single
simulated SeGA detector.

The following effects were taken into account in the simulations:
\begin{itemize}
\item The dependence of the solid angle covered by the detector on the
position of the decay.
\item Attenuation of the $\gamma$-ray flux by the elements of the
plunger and the beam tube.
\item The Doppler shift of the $\gamma$-ray energy depending on the
source velocity and the direction of the $\gamma$-ray emission.
\item The response of the detector to $\gamma$ rays at Doppler shifted
energies (monochromatic sources of radiation were assumed in the
moving reference frame).
\item The angular distribution of $\gamma$-rays in the source
reference frame.
\item The relativistic beaming of $\gamma$-ray in the laboratory
reference frame from the source being in motion (so called ``Lorentz
boost'').  The beaming can be seen from a relation between the
$\gamma$-ray emission angle in the laboratory frame $\theta$ and the
$\gamma$-ray emission angle $\theta'$ in the frame of a nucleus:
\begin{equation}
\cos\theta={{\cos\theta'+\beta}\over{1+\beta\cos\theta'}},
\end{equation}
which result in the following relation
\begin{equation}
d\Omega={{1-\beta^2}\over{(1+\beta\cos\theta')^2}}d\Omega'
={{(1-\beta\cos\theta)^2}\over{1-\beta^2}}d\Omega',
\end{equation}
between the differential solid angles $d \Omega$ and $\d \Omega'$ in
the laboratory and the moving frame of a nucleus, respectively.  The
``Lorentz boost'' results in increased probability of a $\gamma$-ray
emission in the direction of motion of the source despite of the
forward/backward symmetry of the angular distribution in the moving
reference frame.
\end{itemize}

It should be pointed out here that for the above items only the
response of the detector to $\gamma$-rays at non-Doppler shifted
energies, the change of the solid angle and the attenuation of the
$\gamma$-ray flux by the experimental setup can be measured with
calibration sources. All other effects can not be measured directly
and need to be calculated.

The $\gamma$-ray efficiency has been evaluated for the current
experiment assuming three different cases of angular distributions in
the moving reference frame.
\begin{itemize}
\item The angular distribution with coefficients $a_2$=-0.64 and
$a_4$=-0.22 resulting from the intermediate energy Coulomb excitation
calculated for the current experimental conditions according to Ref.\
\cite{Oli04}.

\item The above angular distribution but attenuated due to the
presence of odd-electron charged states to $a_2$=-0.58, $a_4$=-0.13
with the attenuation factor $G_2=4/5$ and $G_4=1/5$ calculated
according to Refs.\ \cite{Stu04} and\ \cite{Gol82}.

\item Isotropic angular distribution.
\end{itemize}

\begin{figure}[h]
\begin{center}
\includegraphics[clip=true,angle=-90,scale=0.5]{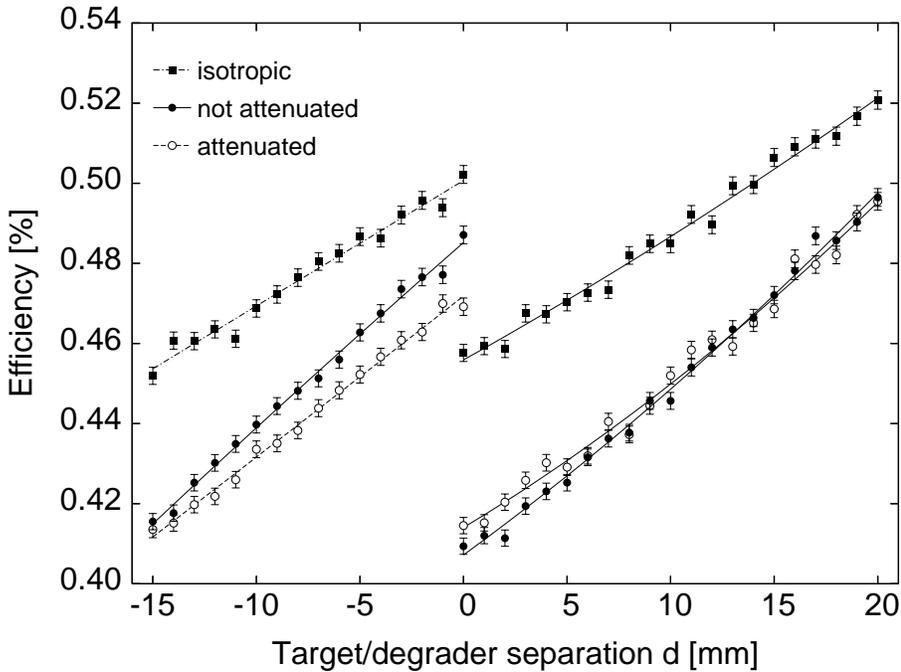}
\caption{Changes in $\gamma$-ray efficiency of a single SeGA detector
at $37^\circ$ as a function of the position of the decay along the
beam axis for the target/degrader separation of $d=15.8$ mm. Filled
boxes/dashed-dotted line, filled circles/continuous line and open
circles/dashed line are for the isotropic, not-attenuated and
attenuated angular distribution in the moving reference frame of the
decaying nuclei, respectively. The discontinuity at $d=0$ mm arises
predominantly due to the change of relativistic beaming after slowing
down in the degrader. For details of the efficiency calculation see
Sec.\ \ref{efficiency} and Tab.\ \ref{efftab}.}
\label{fig3}
\end{center}
\end{figure}

The results of calculations for the target/degrader separation
$d=15.8$ mm are presented in Fig.\ \ref{fig3}.  The discontinuity
observed in this figure at $d=0$ mm arises predominantly from the
change of relativistic beaming after slowing down in the degrader.
The efficiency increase as a function of $z$ is understood from the
fact that the points of decay in the considered range of $z$ are
closer to the detector in the SeGA ring at 37$^\circ$ as $z$
increases. This is true at $\beta \sim 0.3$ for states with lifetimes
on the order of tens of picoseconds which decay within a few
centimeters away from the target.This assumption, however, has to be
verified if longer lifetimes are investigated with the current
method. The trends observed in Fig.\ \ref{fig3} justify the second
order approximation for the efficiency dependence on $z$ employed in
Sec.\ \ref{lifetimeSec}. The efficiency coefficients fitted to the
simulation data are summarized in Tab.\ \ref{efftab} and the resulting
fits are plotted in Fig.\ \ref{fig3}.

\section{Experimental Results}

\subsection{Particle Identification}
\label{PID}

\begin{figure}[h]
\begin{center}
\includegraphics[clip=true,angle=-90, scale=0.50]{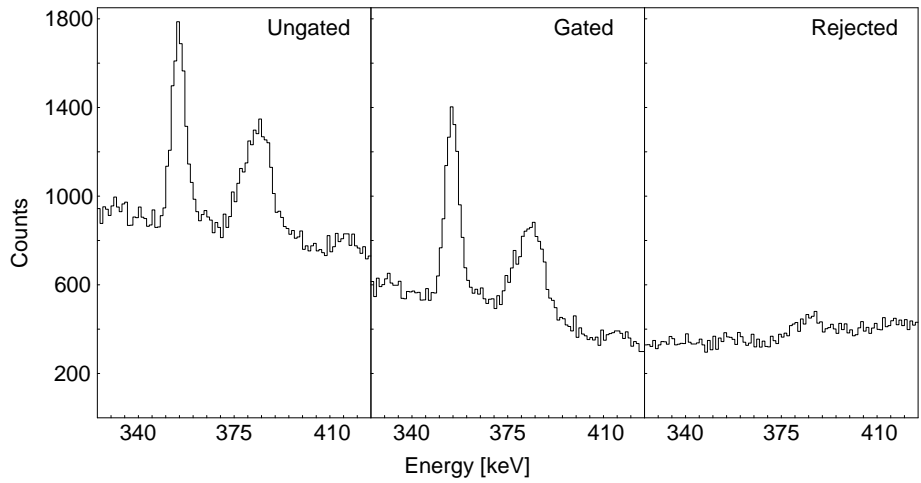}
\caption{Application of the PID gate to a Doppler corrected spectra at
$d$=10.5 mm. The left panel shows the spectrum without the PID
gate. The middle panel shows the same spectrum with the PID gate
applied. The right panel shows events rejected by the PID gate.}
\label{fig4}
\end{center}
\end{figure}

Events that corresponded to the de-excitation process of $^{124}$Xe
were selected using a combination of SeGA\ \cite{Mue01} and detectors
at the focal plane of the S800 spectrometer\
\cite{Baz03,Yur99}. SeGA-S800 time coincidence was measured on an
event-by-event basis starting when the SeGA array detected a
$\gamma$-ray and stopping when the nucleus reached the E1 plastic
scintillator at the focal plane of the S800 downstream from the
degrader.  The $\gamma$-ray energy dependent time gates compensating
for a time-walk resulting from different signal amplitudes in the SeGA
detectors were used off-line to improve the peak to background
ratio. An S800 focal plane gate was created to select events that
corresponded to a $^{124}$Xe nucleus based on the measured energy loss
recorded by the E1 plastic scintillator and the ion chamber. The final
particle identification (PID) gate was a combination of the
time-energy gate and the S800 focal plane gate. The improvement in the
signal to noise ratio from the initial value of $\sim$0.31 in the
region corresponding to the two peaks to the final value of $\sim$0.50
after the application of the particle identification gate can be seen
in Fig. \ref{fig4}.

\subsection{Velocity Measurement}

The precision and accuracy of the $\beta_H$ velocity measurement for
the decays between the target and the degrader has a direct impact on
the precision and accuracy of the lifetime measurement with the
current technique, see Sec. \ref{lifetimeSec}. Below it is argued that
from the Doppler shift of $\gamma$-ray energies emitted in flight, the
error on the velocity measurement $\beta_H$ given by a standard
deviation $\sigma$ can be reduced to less than 0.3\%, and a precise
lifetime can be extracted.

\begin{figure}[hbp]
\begin{center}
\includegraphics[clip=true,angle=-90,scale=.50]{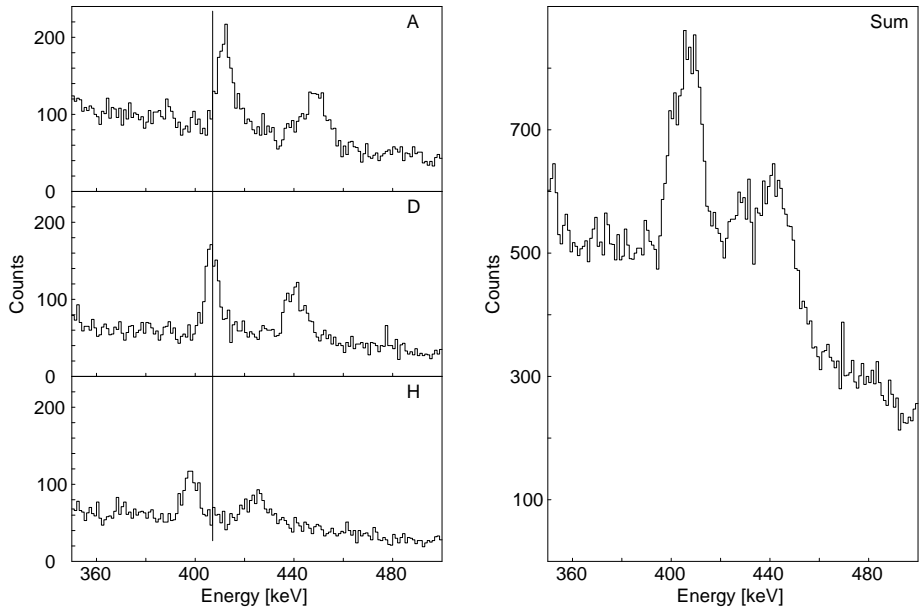}
\caption{Doppler shifted peaks observed at $d$=10.5 mm. The left panel
shows spectra summed over $\varphi$ for $\theta=37^\circ$
corresponding to SeGA detector slices A, D, and H (for naming
conventions of SeGA slices, see Ref. \cite{Mue01}). The right panel
shows the sum of all spectra for the 37$^\circ$ SeGA ring.}
\label{fig5}
\end{center}
\end{figure}

The SeGA detectors are electronically divided into 32 segments which
form 8 slices at different polar angles $\theta$\
\cite{Mue01}. Gamma-ray energies recorded in each of the 8 slices of
individual SeGA detectors were used to measure velocity of the
$^{124}$Xe nuclei through the observed Doppler shift of the $E_0=$354
keV transition depopulating the 2$^{+}$ first excited state. The
non-Doppler corrected spectra corresponding to the SeGA slices and the
PID gate described in Sec. \ref{PID} were sorted for each
target-degrader separation distance. The effect of the Doppler shift
on the lower-energy peak corresponding to $\gamma$ rays emitted after
the degrader can be seen in Fig. \ref{fig5} for a target-degrader
separation $d$=10.5 mm.

\begin{figure}[h]
\begin{center}
\includegraphics[clip=true,angle=-90,scale=0.55]{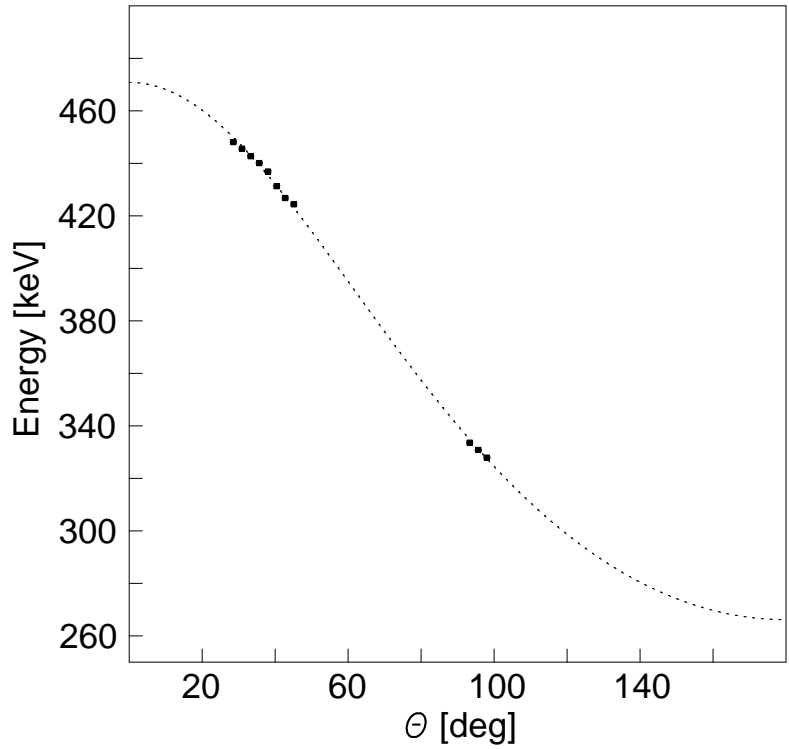}
\caption{The result of a least squares fit of gamma-ray energy
vs. polar angle $\theta$ at $d$=10.5 mm for the peak corresponding to
the decay between the target and the degrader. The data points are the
$\gamma$-ray energies recorded by the SeGA array at different values
of the polar angle $\theta$. The dotted line is the least squares fit
of Eq. \ref{RelDop} to these data points.}
\label{fig6}
\end{center}
\end{figure}

\begin{figure}[h]
\begin{center}
\includegraphics[clip=true,angle=270,scale=0.4]{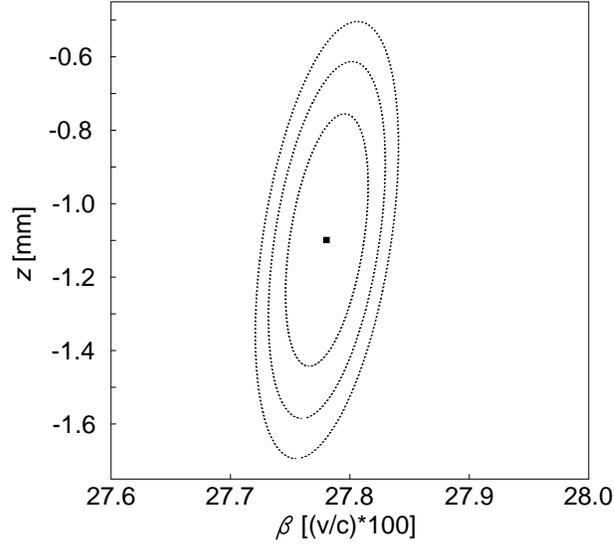}
\caption{The $\chi^2$ contour plot of $\beta_H$ vs. $z_H$ resulting
from the least squares fitting of $\gamma$-ray energy data for
$d$=10.5 mm to Eq.\ref{RelDop}. The dotted ellipses around the
calculated value of $\beta_H$ and $z_H$ correspond to 68.3\%, 95.4\%,
and 99.7\% confidence levels. Notice that the error on $\beta_H$ is
$<0.5\%$ at 99.7\% confidence level. This is strong evidence of the
velocity measurement precision for the time of flight method.}
\label{fig7}
\end{center}
\end{figure}

Spectra corresponding to the de-excitation events recorded by each
slice in the SeGA detectors were summed over the azimuthal angle
$\varphi$ and the centroids of the gamma-ray peaks of interest were
measured at the corresponding value of the polar angle $\theta$. The
centroid data from these spectra were used to determine velocities
$\beta_H$ and $\beta_L$ for the peaks representing $\gamma$-rays
emitted before and after the degrader, respectively, for each
target-degrader separation. These data were fit using the relativistic
Doppler equation:
\begin{equation}
E=E_0{{\sqrt{1-\beta^2}}\over 1-\beta\cos\theta}
\label{RelDop}
\end{equation}
by a least squares fitting method. The average decay position along
the beam axis for the decays upstream and downstream from the
degrader, $z_H$ and $z_L$, respectively, were free parameters of these
fits. The decay position $z$ is related to the polar angle $\theta$ in
Eq. \ref{RelDop} by the target-detector separation distance $L=24$~cm
and the detector segment angle in the array $\theta_0$:
\begin{equation}
\cos\theta={{\cos\theta_0-{z\over L}}\over {\sqrt{1-2({z\over
L})\cos\theta_0+({z\over L})^2}}}.
\label{cos}
\end{equation}
The result of the least squares fit for $\beta_H$ and $z_H$ is shown
in Fig. \ref{fig6} for a target-degrader separation of $d$=10.5
mm. The $\chi^2$ contour plot shown in Fig. \ref{fig7} provides
evidence that velocity $\beta_H$ can be measured with $\sim$0.3\%
precision. The results of the $\beta_H$/$z_H$ and $\beta_L$/$z_L$
measurements are summarized in Tab.\ \ref{velocitytab}. The value of
$\beta_H$ adopted based on the measurements at three target/degrader
separations of $d$=5.3, 10.5 and 15.8 mm is $\beta_H$=0.278(1).

\subsection{Normalized Intensity Measurement}

\begin{figure}[h]
\begin{center}
\includegraphics[clip=true,angle=-90,scale=.4]{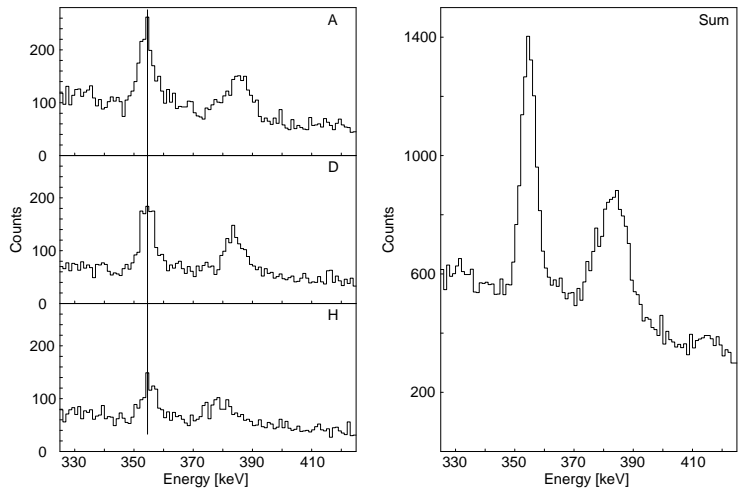}
\caption{Spectra corresponding to those shown in Fig. \ref{fig5} but
Doppler corrected using the $\beta_L$ and $z_L$ values measured for
the decay downstream from the degrader. Notice the improvement in the
signal to noise ratio for the low energy peak over Fig. \ref{fig5}.}
\label{fig8}
\end{center}
\end{figure}

For the intensity measurements as a function of target-degrader
separation, the Doppler corrected spectra at a constant polar angle
$\theta$ were summed over different azimuthal angles
$\varphi$. Doppler corrections were performed separately for the high-
and low-energy peaks with the corresponding measured values of
$\beta_H$ and $z_H$ or $\beta_L$ and $z_L$ respectively. Fig.\
\ref{fig8} demonstrates the improved energy resolution from the
Doppler corrections for the spectrum corresponding to that shown in
Fig.\ \ref{fig5}.

The number of counts in the high- and low-energy peaks were measured
and recorded for each distance $d$. These SeGA photo-peak counts were
normalized to the total number of nuclei detected by the S800 E1
scintillator from the recorded scalers. Since the deadtime of the data
acquisition system was small ($\sim$2\%) and equal for different
target/degrader separations, no further corrections or normalizations
were applied. The resulting normalized intensities are listed in Tab.\
\ref{intensitytab}.

\subsection{Lifetime measurement for the 2$^+_1$ state in $^{124}$Xe}

\begin{figure}[h]
\begin{center}
\includegraphics[clip=true,angle=-90,scale=.5]{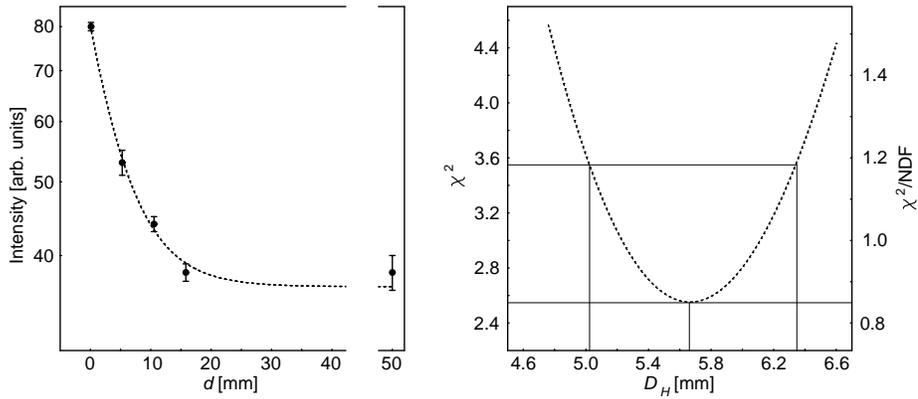}
\caption{Results of an exponent plus a constant fit of Eq.\
\ref{expconst} to the $N_H/N_0$ data from Tab.\
\ref{intensitytab}. The extracted value of $D_H$=5.6$^{+0.8}_{-0.6}$
corresponds to the lifetime $\tau$=64$^{+10}_{-8}$ ps for
$\beta_H$=0.278(1).}
\label{fig9}
\end{center}
\end{figure}

It is indicated by Eq.\ \ref{lowfit} that the lifetime of the 2$^+_1$
state in $^{124}$Xe can be extracted from the decay-curve data in the
last column of Tab.\ \ref{intensitytab} using an exponent plus a
constant fit
\begin{eqnarray}
\label{expconst}
N_L/N_0&=& v\cdot\exp(-{d \over D_H})+u,
\end{eqnarray}
with
\begin{eqnarray}
\label{fitconst}
v&=& {I_0 \over N_0}\tilde{\varepsilon_L}\exp(-{d_0+d_d\over
D_H}),\;\;\;\; u={N_D \over N_0}.
\end{eqnarray}
In the above formulae all the efficiency and background factors are
absorbed into the fitted constants. The results of the fit shown in
Fig.\ \ref{fig9} yield $D_H$=5.6$^{+0.8}_{-0.6}$ mm which corresponds
to $\tau$=64$^{+10}_{-8}$ ps for $\beta_H$=0.278(1) in excellent
agreement with a recent measurement of $\tau$=67.5(1.7) ps reported in
Ref.\ \cite{Sah04}.

The above procedure of extracting lifetime does not take into account
the full experimental information provided by the measurement. A
simultaneous fit to both components of the $\gamma$-ray transition of
interest can be performed; it requires, however, a proper handling of
the efficiency corrections which depend on the target/degrader
separation for the decays upstream from the degrader.

For this simultaneous fit the formulae discussed in Sec.\
\ref{lifetimeSec} can be written as
\begin{eqnarray}
\label{sim}
{N_H\over N_0}&=& p \left[1-\exp(-{d+d_0 \over D_H}\right]+\nonumber\\
&&- q(d+d_0)+ r(d+d_0)(d+d_0+2d_m-2D_H)\nonumber \\ {N_L \over N_0}&=&
w\exp(-{d+d_0 \over D_H})+ u.
\end{eqnarray}
with
\begin{eqnarray}
p= {I_0\over N_0}\tilde{\varepsilon_H},\;\; q= {I_0\over
N_0}\varepsilon_{1\;H},\;\; r= {I_0\over
N_0}\varepsilon_{2\;H},\;\mbox{and}\;\; w= {I_0\over
N_0}\tilde{\varepsilon_L}\exp(-{d_d \over D_H}).
\end{eqnarray}
The total $\chi^2=\chi^2_H+\chi^2_L$ is then minimized as a function
of $D_H$, $d_0$, $d_m$, $p$, $q$, $r$, $u$, and $w$ constants.

In the current analysis the eight-parameter fit implied by Eqs.\
\ref{sim} to nine data points given in Tab.\ \ref{intensitytab} is
carried on as an exercise rather than a proper measurement. In
practical applications additional data points at well chosen
target/degrader separation are needed to constrain the fitted
parameters. Here it was assumed that $d_m=0$ and that the slope of the
$N_H / N_0$ function is zero at $d\sim 15$ mm as suggested by the data
points in Tab.\ \ref{intensitytab}.

\begin{figure}[h]
\begin{center}
\includegraphics[clip=true,angle=-90,scale=.5]{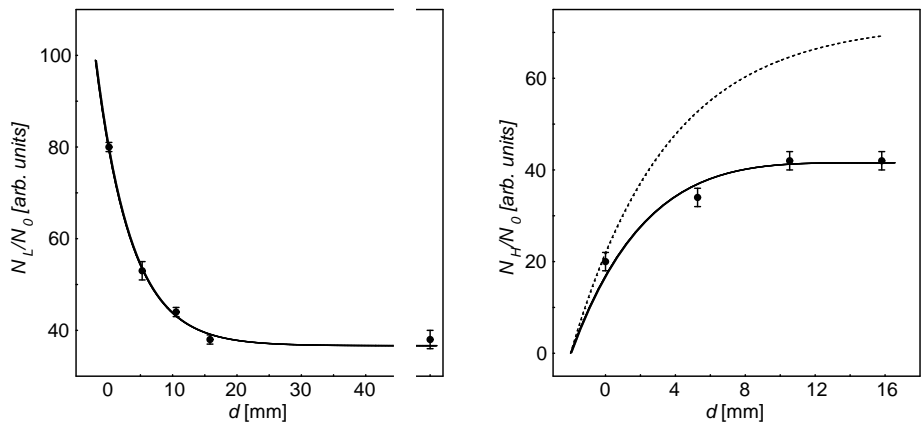}
\caption{Results of a simultaneous fit according to Eqs.\ \ref{sim} to
the data from Tab.\ \ref{intensitytab}. The extracted value of
$D_H$=5.5(7) is consistent with that from the exponent plus a constant
fit from Fig.\ \ref{fig9} to the $N_H/N_0$ data only.  The dashed line
in the right panel corresponds to the curve expected in the absence of
the efficiency loss for the decays between the target and the
degrader.}
\label{fig10}
\end{center}
\end{figure}

The results of the simultaneous fit are sensitive to the $d_0$
parameter through the $N_H/N_0$ component. A fit with the $d_0$
parameter varied freely yields the $D_H$=5.6(6) mm and $d_0$=2.4(4)
mm. This value of $d_0$ seemed too large for the current experimental
conditions. A fit with the $d_0$ parameter constrained to less than
2.0 mm yields $D_H$=5.5(7) mm, a value consistent with that obtained
from the fit to the $N_L/N_0$ data alone. The results of the
simultaneous fit with $d_0$=2.0 mm are shown in Fig.\ \ref{fig10}
together with the $N_H/N_0$ curve expected in the absence of the
$d$-dependent efficiency corrections for the decays between the target
and the degrader. The $w/p \sim0.86$ ratio from this fit is consistent
with the results of the Monte Carlo simulations discussed in Sec.\
\ref{efficiency} and presented in Fig.\ \ref{fig3} and Tab.\
\ref{efftab}.

Figure\ \ref{fig10} indicates that in a plunger experiment with the
stationary degrader the efficiency change as a function of the
target/degrader separation has a significant impact on the detected
number of counts for the decays upstream from the degrader. The
efficiency corrections for the decays upstream from the degrader can
be measured if a sufficient number of data points at $d>D_H$ is
provided. On the other hand the $N_L/N_0$ data are essentially free
from the efficiency corrections for a large range of lifetimes. These
factors are of significant importance and need to be taken into
account in planning and executing relativistic plunger experiments.

\section{Range of application for the method}

The factor which is limiting the application of the method for short
lifetimes is related predominantly to the thickness of the plunger
excitation target and the degrader foil. Certainly, the lifetime of a
state being studied has to be long enough for the excited nuclei to
emerge from the target and decay in vacuum. A typical target thickness
for experiments with relativistic beams is $\sim$ 100 $\mu$m which
corresponds to $\sim$1 ps at $\beta \sim 0.3$.

If thin excitation targets can be used in an experiment the $d_0 <
D_H$ condition sets the limit for observing variations in $N_H$ as a
function of $d$ according to Eq.\ \ref{NH}. The minimization of $d_0$
is a well known challenge in plunger experiments; it requires that the
target and the degrader foils are flat and parallel. This is achieved
by careful stretching and alignment of the target and degrader foils.
Currently the values of $d_0<10$ $\mu$m, corresponding to 0.3 ps at
$\beta=0.3$, are routinely achieved. However, it must be stressed that
the targets of $\sim$10 $\mu$m thickness are essential for
sub-picosecond lifetime measurements and that application of these
targets necessarily results in reduced reaction yields.

The shortest lifetime which can be measured from the variation of
$N_L$ is defined by the degrader thickness entering the $d_0+d_d<D_H$
condition from Eq.\ \ref{NL}. The degrader thickness is selected to
provide a change of the Doppler shift which is sufficient to resolve
the components of the transition of interest emitted before and after
traversing the degrader.  For typical $\gamma$-ray detection
resolution in relativistic beam experiments of $\sim$2\% FWHM the
degrader thickness of $\sim$300 $\mu$m is required.  This sets the
lifetime limit of more than 3 ps at $\beta\sim 0.3$. However, for such
short lifetimes the decays at changing velocity within the degrader
contribute significantly and peak shape modeling based on the stopping
power of the degrader seems to be needed to account properly for the
intensity balances. The contribution of the decays in the degrader
decreases for longer lifetimes and seem to be negligible in the
current analysis as demonstrated from experimental spectra shown in
Fig.\ \ref{fig8}.

For the short lifetime measurements discussed above the efficiency
changes as a function of the target/degrader separation are
essentially negligible. However, these efficiency changes limit the
application of the method for states with long lifetimes. It is
estimated that the forward ring of the SeGA array at 37$^\circ$ can be
used to detect decays up to 20 cm downstream from the degrader
foil. For a significant fraction of the decays of interest to occur
with this length the $D_H$ has to be shorter by a factor of
$\sim$4. This corresponds to the lifetime of $\tau\sim 500$ ps.

Based on the above discussion, the current technique can be applied to
$\gamma$-ray emitting states with lifetimes between 5 and 500 ps. The
above estimates do not take into account the relativistic time
dilation which is a $\sim$10\% correction at $\beta \sim 0.3$ which
allows to investigate shorter lifetimes. It seems possible to extend
the method to sub-picosecond lifetime measurement, however, further
careful studies are needed for such applications.

\section{Choice of the relativistic plunger target/degrader materials}

An important issue for the application of the current technique is a
choice of the target and degrader materials. The thickness of the
degrader is directly related to the constant contribution $N_D$ to the
intensity $N_L$ used to determine the lifetime from Eq.\ \ref{lowfit}
and\ \ref{expconst}. This arises from the fact that relativistic beams
have enough energy to react on both, the plunger excitation target and
the degrader. This is one of the essential differences between the
current method and plunger applications near the Coulomb barrier for
which reactions on the degrader can be, for many cases, completely
eliminated by a proper choice of the degrader material. The precision
of the $N_D$ determination has a direct impact on the precision of the
lifetime measurement. The precision of the lifetime measurement can be
improved by minimizing the $N_D$, however, it is not essential for the
application of the method to eliminate $N_D$ completely.

The current approach for selecting the target and degrader material
for relativistic plunger experiments is based on two principles:
\begin{itemize}

\item The cross section for the excitation process of interest has to
be maximized on the target and minimized on the degrader. For example,
if Coulomb excitations are used, the target and the degrader should be
made out of high-Z and low-Z material, respectively. On the other
hand, if fragmentation is used to populate the states of interest then
the target and the degrader should be made out of low-Z and high-Z
material, respectively.

\item The thickness of the degrader has to be minimized preserving the
change of velocity sufficient to resolve the Doppler-shifted
components of the transition of interest emitted before and after
traversing the degrader. Thus, the change of the Doppler shift from
traversing the degrader has to be balanced against the $\gamma$-ray
energy resolution for the peaks of interest. Since the energy
resolution is determined by various factors\ \cite{Gla98} including
solid angle for $\gamma$-ray detection and beam momentum spread, case
by case studies are needed for planning a relativistic plunger
experiment.

\end{itemize}

\section{Conclusions}

The experiment reported in the current paper provides a proof of
principle for application of the time of flight technique for lifetime
measurements with relativistic beams of heavy nuclei. The
$\tau$=64$^{+10}_{-8}$ ps lifetime for the first excited $2^+_1$ state
in $^{124}$Xe measured with the above relativistic plunger method is
in an excellent agreement with the most accurately known value of
$\tau$=67.5(1.7) ps from recent Ref.\ \cite{Sah04}. Further
development of the method for experiments with rare isotopes produced
in projectile fragmentation reaction at the NSCL are planned. In
particular a plunger device optimized for the fast beams at the NSCL
is being constructed by the collaboration between the NSCL and the
University of Cologne.

The advantage of the plunger method is in its relatively simple
dependence on reaction kinematics and in the fact that it does not
depend on reaction dynamics. Thus, in a plunger experiment, any
excitation mechanism can be used for populating the state of interest
while the lifetime information is extracted from a decay curve
measured directly. Moreover, the lifetime information results from a
series of measurements for which only relative properties of the
experimental set-up need to be known or calibrated. In particular, the
knowledge of the absolute set-up efficiency is not critical for
accurate lifetime measurement.

The plunger method is not limited by isomer contamination of the beam.
This makes it an ideal tool to explore odd-even and odd-odd nuclei,
investigating in particular single-particle electromagnetic moments
near closed shells in neutron rich nuclei for which the available
information is scarce at present. It is estimated that plunger
experiments require beam intensities of only 10$^6$ particle per
second, so that knockout reactions with rare isotopes are possible.

Current applications of the plunger technique near the Coulomb barrier
provide the most accurate electromagnetic transition rate measurements
for picosecond lifetimes. Successful application of the relativistic
plunger technique in fragmentation facilities provides an opportunity
to explore new and untouched areas of rare isotopes far from
stability.

\section{Acknowledgments}

This work has been partially supported by the (US) National Science
Foundation Grant No. PHY01-10253 and by the BMBF (GERMANY) under
contract No. 06K167.

\newpage

\begin{table}
\caption{Coefficients for $\gamma$-ray efficiency $\varepsilon(z)$
dependence on the position of the point of decay $z$ along the beam
axis fitted according to Eq.\ \ref{parabola} to the simulation data
shown in Fig\ \ref{fig3}. See Sec.\ \ref{efficiency} for details.}
\begin{tabular}{||c||c|c|c||c|c|c||}
\hline\hline angular distribution & $\varepsilon_{0\;H}$ &
$\varepsilon_{1\;H}$ &$\varepsilon_{2\;H}$ &$\varepsilon_{0\;L}$
&$\varepsilon_{1\;L}$ &$\varepsilon_{2\;L}$ \\ \hline Coulex &
0.4852(5) & 0.00451(6) & -0.000011(5) & 0.4072(5) & 0.00376(4) &
0.000038(3)\\ attenuated Coulex & 0.4718(5) & 0.00404(6) & 0.000002(5)
& 0.4140(5) & 0.00310(4) & 0.000048(3) \\ isotropic & 0.5007(6) &
0.00313(6) & 0.000000(5) & 0.4559(5) & 0.00289(4) & 0.000019(3) \\
\hline\hline
\end{tabular}
\label{efftab}
\end{table}

\begin{table}
\caption{Result of $\beta_H$/$z_H$ and $\beta_L$/$z_L$ measurements as
a function of the target/degrader separation $d$.}
\begin{tabular}{||r||c|c||c|c||}
\hline\hline $d$ & $\beta_H$ & $z_H$ [mm] & $\beta_L$ & $z_L$ [mm]\\
\hline 0.0 & & & 0.1784(2) & 7.6(3) \\ 5.3 & 0.2771(5) & 1.3(5) &
0.1784(3) & 7.9(4) \\ 10.5 & 0.2778(3) & -1.1(3) & 0.1792(3) & 8.3(4)
\\ 15.8 & 0.2774(5) & -6.7(5) & 0.1792(4) & 8.3(5) \\ $\infty$ & & &
0.1779(4) & 7.2(7) \\ \hline\hline
\end{tabular}
\label{velocitytab}
\end{table}

\begin{table}
\caption{Normalized intensities for the high- and low-energy component
of the $E_0=354$ keV $2^+_1 \rightarrow 0^+_1$ transition in
$^{124}$Xe measured in the current experiment as a function of the
target/degrader separation $d$.}
\begin{tabular}{||r||c||c||}
\hline\hline $d$ & $N_H$/$N_0$ & $N_L$/$N_0$ \\ \hline 0.0 & 20(2)
&80(1) \\ 5.3 & 34(2) & 53(2) \\ 10.5 & 42(2)& 44(1) \\ 15.8 &42(2) &
38(1) \\ $\infty$ & & 38(2) \\ \hline\hline
\end{tabular}
\label{intensitytab}
\end{table}

\newpage

\end{document}